\documentclass[aps,pra,twocolumn,superscriptaddress,preprintnumbers,times]{revtex4}
\usepackage{epsfig,color,braket}
\usepackage{graphicx,amssymb,epsfig,amsmath,amsfonts,amssymb,mathtools}

\newcommand{\be}{\begin{eqnarray}}
\newcommand{\ee}{\end{eqnarray}}

\DeclareMathOperator{\Tr}{Tr}

\begin{document}

\title{Testing dimension and non-classicality in communication networks}

\author{Joseph Bowles}
\affiliation{D\'epartement de Physique Th\'eorique, Universit\'e de Gen\`eve, 1211 Gen\`eve, Switzerland}
\author{Nicolas Brunner}
\affiliation{D\'epartement de Physique Th\'eorique, Universit\'e de Gen\`eve, 1211 Gen\`eve, Switzerland}
\author{Marcin Paw\l{}owski}
\affiliation{Institute of Theoretical Physics and Astrophysics, University of Gda\'nsk, 80-952 Gda\'nsk, Poland}

\begin{abstract}
We consider networks featuring preparation, transformation, and measurement devices, in which devices exchange communication via mediating physical systems. We investigate the problem of testing the dimension of the mediating systems in the device-independent scenario, that is, based on observable data alone. A general framework for tackling this problem is presented, considering both classical and quantum systems. These methods can then also be used to certify the non-classicality of the mediating systems, given an upper bound on their dimension. Several case studies are reported, which illustrate the relevance of the framework. These examples also show that, for fixed dimension, quantum systems largely outperform classical ones. Moreover, the use of a transformation device considerably improves noise tolerance when compared to simple prepare-and-measure networks. These results suggest that the classical simulation of quantum systems becomes costly in terms of dimension, even for simple networks.
\end{abstract}

\maketitle

\section{Introduction}
The problem of estimating the dimension of an unknown physical system has attracted attention recently. Following early works discussing the problem in the context of Bell inequalities \cite{brunner08,vertesi08,david08}, a framework was presented for the simplest case of a prepare-and-measure scenario \cite{DimWit}. Such a setup features two devices. First a preparation device, which allows the observers to prepare a physical system in various ways. Second, a measurement device, which allows the observer to perform a measurement on the prepared physical system. It is then possible to find the minimal dimension of the physical system that is compatible with the data. The method is device-independent (DI), in the sense that dimension can be certified from the data alone. Techniques tailored for classical \cite{DimWit,arno1}, and quantum \cite{wehner,BNV,NVdim} systems were reported, as well as for the case in which the devices are assumed to be independent \cite{Bowles,arno2}. The practical relevance of these ideas was recently illustrated \cite{hendrych,ahrens}. Also, the notion of dimension was discussed in more general models beyond quantum theory \cite{dimGPT}.

A closely related problem is that of testing the non-classicality of communication. More specifically, considering again the prepare-and-measure setup, it is possible to guarantee the use of quantum communication, under the assumption that the dimension of the system is upper bounded \cite{DimWit}. From a conceptual point of view, this approach aims at quantifying how much classical communication is required to simulate quantum communication \cite{galvao,harrigan}, a relevant problem in the foundations of quantum theory and in communication complexity \cite{harry_review}. Moreover, these ideas are relevant for `semi-device-independent' quantum information processing \cite{PB11}. Here the correct implementation of a protocol can be guaranteed in a device-independent way, with an additional assumption on the Hilbert space dimension. Protocols for semi-DI quantum key distribution \cite{PB11,Woodhead1,Woodhead}, randomness certification \cite{Letal11,HWL12}, and the characterization of quantum systems \cite{YCL,VN11} were discussed, with experimental implementations recently reported \cite{lunghi15,marcin14,bennet}.

\begin{figure}[t!]
\includegraphics[scale=0.6,keepaspectratio=true]{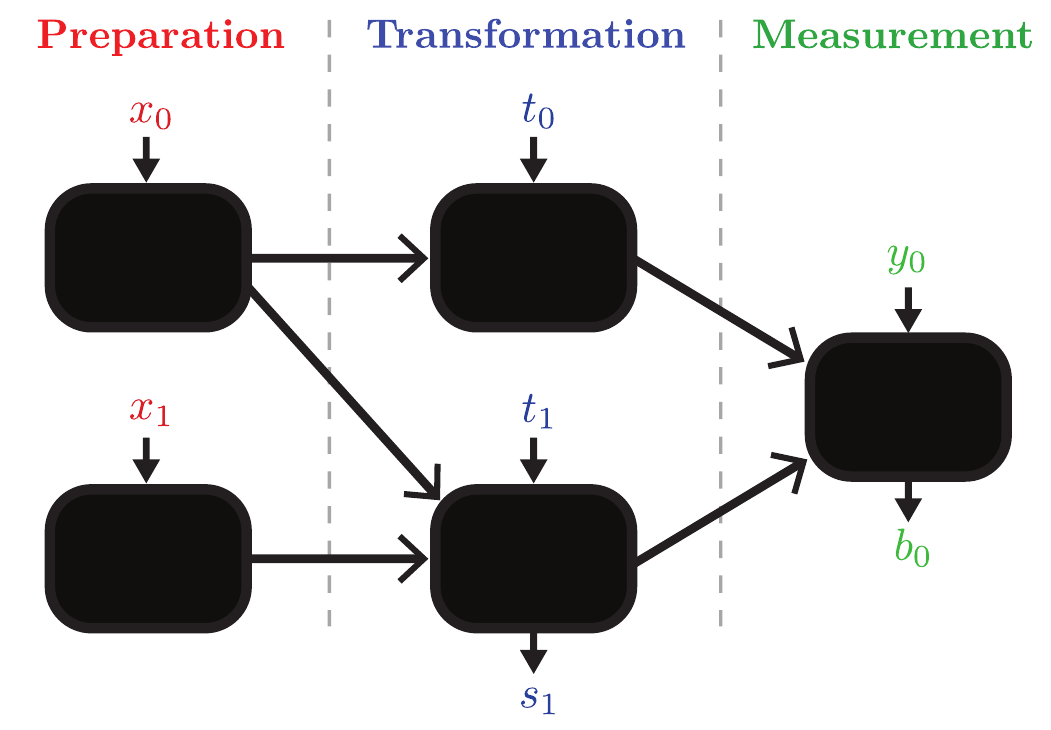}
\caption{\label{fig1} We consider networks featuring preparation, transformation and measurement devices. All devices receive classical inputs. Transformation and measurement devices provide classical outputs. The arrows between the devices represent communication channels, either quantum or classical.}
\end{figure}

More generally, it is natural to consider the problem of testing dimension and non-classicality in general communication networks, in which black-box devices exchange and process information. To model such a situation, we consider a network composed of preparation devices, transformation devices, and measurement devices (see Fig.1). First, the preparation devices send out information encoded in physical systems of certain dimension. In turn, these physical systems (and the information they carry) are processed in transformation devices. Finally, the systems are measured (i.e. the information is extracted) using measurement devices. Since we work in the device-independent picture, all devices are represented by black boxes. We therefore have access only to measurement data, that is the probabilities of obtaining certain measurement results, given the choices of preparations, transformations, and measurements made by the observer. From this data, our goal is then to infer a lower bound on the dimension of the physical systems mediating the information. We will here consider both the case of classical and quantum systems. Moreover, we discuss testing the non-classicality of communication under the assumption that the dimension is upper bounded. Note that the definition of dimension that we employ here is related to the number of perfectly distinguishable states, i.e. that there should be precisely $d$ perfectly distinguishable states in dimension $d$. For classical and quantum systems this will coincide with the classical alphabet size and Hilbert space dimension respectively.

We start by describing the general scenario we consider in Section \ref{general scenario}. Next, we discuss a general framework for addressing this problem for the case of classical systems (Section \ref{classical networks}) and quantum systems (Section \ref{quantum networks}). For the sake of clarity, we present the framework in detail for a simple network, featuring one preparation, one transformation, and one measurement device. We show that the idea of dimension witnesses \cite{DimWit} can be generalized to arbitrary networks, and present methods for deriving optimal witnesses. In Section \ref{nonclassicality}, we show how dimension witnesses can be used to certify and measure non-classicality of communication. In order to illustrate the relevance of these methods, we discuss several case studies in Section \ref{case studies}, deriving and characterizing dimension witnesses for simple networks. An interesting feature shared by most of these examples is the fact that quantum systems strongly outperform classical systems of the same dimension. In fact, we observe a significant enhancement of the advantage offered by quantum systems over classical ones compared to the usual prepare-and-measure scenario. This suggests interesting possibilities for quantum information protocols, and for addressing questions in the foundations of quantum theory. These issues are discussed at the end of the paper, in Section \ref{discussion}.

\section{General Scenario}\label{general scenario}
The general scenario we wish to consider is a network of devices exchanging and processing information, as represented in Fig. \ref{fig1}. Devices are represented by black boxes. An arrow connecting two devices represents a (one-way) communication channel between them \footnote{The case of two-way communication could also be considered, but we will not discuss it here.}.

A network consists of three levels: (i) a number of preparation devices, (ii) a number of transformation devices and (iii) a number of measurement devices. In each round of the experiment, the observer chooses the preparations $\bf{x}$, the transformations $\bf{t}$ and the measurement settings $\bf{y}$. He then obtains measurement outcomes $\bf{b}$; note that transformation devices can also provide outcomes, denoted $\bf{s}$. More precisely, we have that the choice of preparations is
given by ${\bf{x}}= \{ x_i\}$, where $x_i$ denotes the input for device $i$. The choice of transformations is ${\bf{t}}= \{ t_j\}$, where $t_j$ denotes the input for device $j$, and the (possible) outcomes are ${\bf{s}}= \{ s_j\}$, where $s_j$ denotes the output of device $j$. Finally, the choice of measurement settings is ${\bf{y}}= \{ y_k\}$, where $y_k$ denotes the input for measurement device $k$, and gives outcomes ${\bf{b}}= \{ b_k\}$, where $b_k$ is the output of measurement device $k$. The experiment is therefore characterized by the data
\be \label{data} p({\bf{b}},{\bf{s}}|{\bf{x}},{\bf{t}},{\bf{y}}), \ee
that is, the conditional probabilities of observing outputs ${\bf{b}},{\bf{s}}$ given inputs ${\bf{x}},{\bf{t}},{\bf{y}}$. A general scenario is thus specified by a directed graph representing the network, and the number of inputs and outputs for each of the devices (which we will here consider to be finite).

In this network, the devices exchange information encoded in physical systems. For instance, upon receiving input $x_i$, each preparation device emits a system, the state of which is adapted depending on $x_i$. Which physical system is used, and what mechanism is used to encode information in it, is completely unknown to the observer, who has only access to inputs and outputs of the black boxes. That is, we work in a device-independent scenario.

Now the main point is the following. Clearly, the amount of information about $x_i$ which can be encoded in the system will depend on its dimension (i.e. the number of independent degrees of freedom of the system). Therefore, we expect that a restriction on the dimension will in general limit the possible observable data \eqref{data}. Consider for instance the case in which the outputs ${\bf{b}}$ contain all information about the inputs ${\bf{x}}$. This implies that the mediating physical systems had enough dimensions for encoding ${\bf{x}}$ perfectly.

The main question we will discuss in the present work is to understand the limitations on the data, arising from constraints on the dimension of the mediating systems. This will allow us to find lower bounds on the dimension of the systems present in a network for given data \eqref{data}. In particular, we will discuss bounds for both classical and quantum systems. Notably, we will see that for a fixed dimension, quantum systems outperform classical ones.

\section{Classical networks}\label{classical networks}

\begin{figure}[t]
\includegraphics[scale=0.6,keepaspectratio=true]{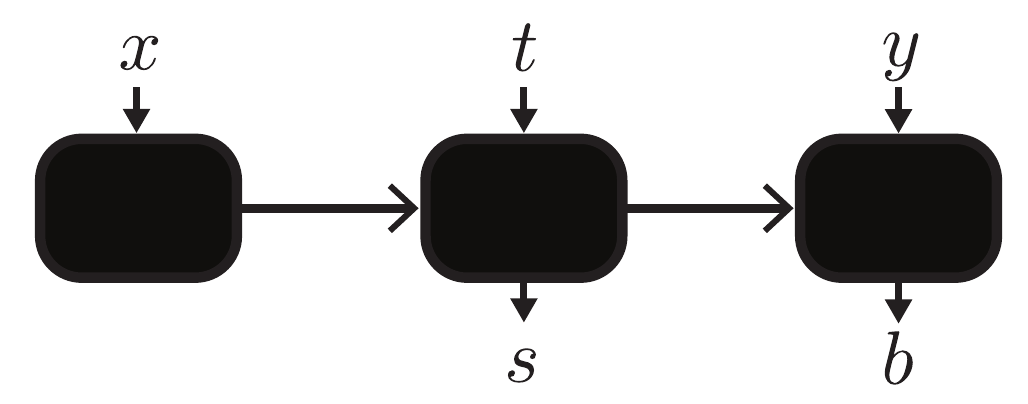}
\caption{\label{line} A simple network consisting of a preparation, a transformation and a measurement device. The set of possible distributions of inputs and outputs, $p(bs|xty)$, will depend on the dimension of the communication allowed between the devices, and whether the communication is classical or quantum.}
\end{figure}

For the sake of clarity, we will focus on the network consisting of one preparation device, followed by a single transformation device, and finally a single measurement device (see Fig. \ref{line}). The data is thus given by the conditional distribution $p(b,s|x,t,y)$; we consider a finite (but otherwise unspecified) number of inputs and outputs. Note that the methods discussed below can be straightforwardly generalized to more general networks.

\subsection{Basics}

We start our analysis by considering classical communication between the devices. Denote by $c_0$ the communication sent from the preparation device to the transformation device, and $c_1$ the communication sent from the transformation device to the measurement device. We consider communication of bounded dimension $d$, that is

\be c_0, c_1 \in \{  1, \cdots ,d \}. \ee
Upon receiving input $x$, the preparation device sends communication $c_0$, with probability $p(c_0|x)$. In turn, upon receiving input $t$ and communication $c_0$ (from the preparation device), the transformation device outputs $s$ and sends communication $c_1$ to the measurement device with probability $p(s,c_1|t,c_0)$. Finally, upon receiving measurement setting $y$ and communication $c_1$, the measurement device outputs $b$ with probability $p(b|y,c_1)$. We thus have that
\begin{align}\label{classicaldet}
p(b,s|x,t,y)=\sum_{c_0,c_1=1}^{d}p(c_0|x)p(s,c_1|t,c_0)p(b|y,c_1).
\end{align}

We first consider the case in which all devices act deterministically. That is, each of the previously mentioned probabilities are either $0$ or $1$. It follows that each probability $p(b,s|x,t,y)$ also takes only values $0$ or $1$. We refer to these sets of data as `deterministic strategies'.

In general, we also want to include the possibility that the devices in the network output probabilistically, and moreover that they follow a common strategy. That is, the behaviour of the devices might be correlated, due to some (common) internal variable $\lambda$ (referred to as shared randomness). The set of possible distributions now becomes all convex combinations of deterministic strategies:
\begin{align}\label{cprobs}
&p(b,s|x,t,y)=\\
&\;\;\int_\lambda\pi(\lambda)\text{d}\lambda\sum_{c_0,c_1=1}^{d}p_{\lambda}(c_0|x)p_{\lambda}(s,c_1|t,c_0)p_{\lambda}(b|y,c_1) \nonumber,
\end{align}
where $\pi(\lambda)$ is a normalized probability density over $\lambda$ and $p_{\lambda}(c_0|x)$ denotes the probability for the preparation device to send $c_0$, given input $x$ and internal variable $\lambda$, and so on.

Any set of data that cannot be decomposed in the form \eqref{cprobs} therefore requires the use of communication ($c_0$ and/or $c_1$) of dimension strictly greater than $d$. In the next sections we will see how to test whether a given set of data can be decomposed in the above form or not. This will provide the `dimension witnesses' we are looking for.

\subsection{Geometrical interpretation}\label{geo}

The above ideas admit an elegant description in geometrical terms. Initially developed in the context of Bell nonlocality \cite{brunner_review}, these ideas were also adapted to the prepare-and-measure scenario \cite{DimWit}.

The goal here is to characterize the set of distributions \eqref{cprobs} in geometrical terms. Consider first one particular set of data $p(b,s|x,t,y)$. This distribution can be viewed as a vector ${\bf p}$ where each component of the vector corresponds to one of the probabilities $p(b,s|x,t,y)$ appearing in the data. Hence ${\bf p} \in \mathbb{R}^{D}$, where
\be D= |b| \, |s| \, |x| \, |t| \,|y| \ee
with $| b |$ denoting the alphabet size of $b$, that is the number of possible outcomes $b$, and similarly for other symbols.

Next, consider the entire set of distributions admitting a decomposition of the form \eqref{cprobs}, that is, all sets of data that can be obtained by using communication $c_0$ and $c_1$ of dimension $d$. This set, denoted $\mathbb{P}_{d}$, thus forms a subspace of
$ \mathbb{R}^{D}$. In fact, $\mathbb{P}_{d}$ forms a convex polytope. Its extremal points (or vertices) correspond to the deterministic strategies, that is, the set of distributions of the form \eqref{classicaldet}, for which $p(b,s|x,t,y) \in \{0,1 \}$ for all $b,s,x,t,y$. Alternatively, the polytope $\mathbb{P}_d$ can also be characterized by its facets (of which there is a finite number, since the number of vertices is finite). Formally, facets are given by linear inequalities
\be {\bf p}\cdot {\bf A} = \sum_{b,s,x,t,y} \alpha^{b,s}_{x,t,y} \, p(b,s|x,t,y) \leq C_d
\ee
where $ \alpha^{b,s}_{x,t,y} $ and $C_d$ are real numbers (usually integers). ${\bf A}$ is the $D$-dimensional vector, with components $\alpha^{b,s}_{x,t,y}$, associated to the facet, i.e. orthogonal to the hyperplane given by the facet. Therefore we have that
\begin{align}
{\bf p} \in\mathbb{P}_{d} \iff {\bf p} \cdot {\bf A} \leq C_d
\end{align}
where the right-hand side means that all facet inequalities are satisfied. Moreover, we have that $\mathbb{P}_{d}\subseteq \mathbb{P}_{d+1}$, since all strategies involving $d$-dimensional communication can always be realized using communication of dimension $d+1$.

In practice, the polytope $\mathbb{P}_{d}$ can be constructed for simple networks, i.e. few devices and small alphabets for the inputs and outputs. Specifically, one starts by listing the deterministic strategies, i.e. the vertices of the polytope. Then, appropriate software (see e.g. \cite{porta,lrs}) allows one to find the facets of the polytope. Beyond simple cases however, the problem becomes  intractable on standard computers.

Finally, note that one can slightly reduce the complexity of the problem by taking into account certain constraints on the data $p(b,s|x,t,y)$. This allows one to discard certain (redundant) components of $\vec{p}$. In particular, we have here the normalization conditions
\be \sum_{b,s} p(b,s|x,t,y) = 1   \quad \forall x,t,y \ee
and the condition that
\be \sum_{b}  p(b,s|x,t,y) = p(s|x,t)   \quad \forall s,x,t,y. \ee
That is, the output $s$ of the transformation device does not depend on the choice of input $y$ for the measuring device. This follows from the fact that $y$ can in principle be chosen after the output $s$ is obtained. For more general networks, it is important to take all such `no-signaling' conditions into account in order to reduce the complexity of the problem.

\subsection{Classical dimension witnesses}

Our main goal is to develop methods for testing whether a given set of data $p(b,s|x,t,y)$ is compatible with a particular network sending communication of bounded dimension. To address this question, we will now discuss the concept of `dimension witnesses', hence generalizing the ideas of Ref. \cite{DimWit} to networks.

Consider linear combinations of the form:
\begin{align}\label{claswit}
W= {\bf w} \cdot {\bf p}=  \sum_{ b,s,x,t,y}  \omega^{ bs}_{xty} p(b,s|x,t,y) \leq C_d,
\end{align}
where {\bf w} is a $D$-dimensional vector, with real components $\omega^{ bs}_{xty}$, and $C_d$ is a real number. We say that an inequality of the above form is a \emph{linear classical dimension witness of dimension $d$}, if (i) the inequality holds for any distribution $p(b,s|x,t,y)$ realizable with communication of dimension $d$, and (ii) there exists at least one distribution $p(b,s|x,t,y)$ (involving systems of dimension at least $d+1$) for which the inequality is violated.

The geometrical ideas discussed in the previous subsection are relevant here, as they will allow us to construct dimension witnesses. Take one facet inequality of the polytope $\mathbb{P}_{d}$ : property (i) above will immediately be satisfied. In general, there will also exist a vector ${\bf p}\in\mathbb{P}_{d'}$ with $d<d'$ that will violate the facet inequality, and hence (ii) is also satisfied. Such facet inequalities will be called `tight dimension witnesses'. In fact, the complete list of the facets of $\mathbb{P}_{d}$ will provide a complete list of dimension witnesses, which allow one to find the minimal dimension of the communication necessary to reproduce a given set of data.

In the section \ref{case studies}, we will present several examples of dimension witnesses.

\section{Quantum networks} \label{quantum networks}

We now move to the case of quantum communication networks. Here, the classical channels are replaced by quantum channels. Our goal is thus to characterize the sets of data compatible with sending quantum communication of bounded Hilbert space dimension in the network. For the sake of clarity, we will also focus on the simple network of Fig. \ref{line}.

\subsection{Basics}

Consider again the network consisting of one preparation device, followed by a transformation device, and finally by a measurement device. The devices can now produce, process, and measure quantum systems. The constraint we consider is that the quantum systems transmitting information between the devices are of Hilbert space dimension bounded by $d$. 

Let us first consider the preparation device. Upon receiving input $x$, the device prepares a $d$-dimensional quantum system in state $\rho_{x}$, which is sent to the transformation device. In turn, the transformation device receives input $t$, as well as the quantum communication $\rho_{x}$,  produces an outcome $s$, and sends a $d$-dimensional quantum system to the measurement device. The action of the transformation device can thus be represented by a set of completely positive (CP) maps $\{ \Phi_{s|t}\}$ (acting on $\mathbb{C}^d$), such that $\sum_{s} \Phi_{s|t}$ is completely positive and trace preserving (CPTP): this ensures that $\sum_s p(s|x,t)=1$ for all $x,t$. Note that, since we impose that all communication is of bounded dimension $d$, we restrict to CP maps which do not increase the Hilbert space dimension \footnote{Indeed, more general transformations, which increase the Hilbert space dimension, could be considered.}. With probability $\Tr[\Phi_{s|t}(\rho_x)]$ the transformation device outputs $s$, and sends the quantum state
\be \Phi_{s|t}(\rho_x)/\Tr[\Phi_{s|t}(\rho_x)] \ee
to the measuring device. Finally, upon receiving this quantum communication and the input $y$, the measuring device provides an output $b$. This is represented by a set of measurement operators $M_{b|y}$ (acting on $\mathbb{C}^d$), such that $M_{b|y}\geq 0$ and $\sum_b M_{b|y} = \mathbb{I}$.

Putting all this together we obtain that
\begin{align}\label{qprobdet}
p(b,s|x,t,y)=\Tr \left( \Phi_{s|t}(\rho_x)M_{b|y}\right).
\end{align}
Any set of data admitting a decomposition of this form is thus realizable with quantum communication of dimension $d$. On the contrary, if such a decomposition cannot be found, then higher dimensional quantum systems must have been used.

As in the case of classical networks, it is also relevant to allow for the devices to act according to a common strategy $\lambda$. In this case, the set of compatible distributions is therefore the convex hull of those of the form \eqref{qprobdet}:
\begin{align}\label{qdist}
p(b,s|x,t,y)=\int_\lambda\Tr \left( \Phi^{\lambda}_{s|t}(\rho^{\lambda}_x)M^{\lambda}_{b|y}\right)\pi(\lambda)\text{d}\lambda,
\end{align}
where now the states, transformations and measurements are written with $\lambda$ dependence. Finally, note that one could also consider the case in which the devices share quantum correlations, i.e. initial entanglement (see Section \ref{case4} for an example).

\subsection{Quantum dimension witnesses}

The problem is now to test whether a given set of data $p(b,s|x,t,y)$ is compatible with a particular network sending quantum communication of bounded Hilbert space dimension. Similarly to the classical case discussed above, we now define `quantum dimension witnesses'.

Consider again linear inequalities of the form
\begin{align}\label{qwit}
W={\bf w} \cdot {\bf p}= \sum_{b,s,x,t,y}\omega^{bs}_{xty} \, p(b,s|x,t,y) \leq Q_{d},
\end{align}
with ${\bf w} $ a $D$-dimensional vector, with real components $\omega^{bs}_{xty}$, and $Q_d $ a real number. In analogy to the classical case, $W$ is a \emph{linear quantum dimension witness of dimension $d$} if (i) the above inequality is satisfied by all sets of data $p(b,s|x,t,y)$ realizable with quantum communication of dimension $d$, and (ii) using quantum communication of dimension greater than $d$ allows one to violate the inequality.

Finding quantum dimension witnesses is generally a harder task than in the classical case. To the best of our knowledge, there are no known efficient computational methods for this problem; see however Refs \cite{NVdim} for recent progress.

\section{Testing non-classicality}\label{nonclassicality}

An interesting development related to dimension tests is the possibility of certifying non-classicality of communication in a device-independent way, assuming an upper-bound on the dimension. This aspect was discussed in Ref. \cite{DimWit} for simple prepare-and-measure scenarios. Here we consider this problem in the context of more general networks.

Before moving on, it is important to understand why an assumption on the dimension is necessary in order to make the problem non-trivial. Consider for instance the network of Fig. 2. If the dimension is not limited, then the input settings of the preparation and transformation devices, $x$ and $t$, can be perfectly transmitted to the final measurement device. Since the transformation device has all information about $x$ and $t$, and the measuring device has all information about $x,t,y$, it follows that any possible statistics $p(b,s|x,t,y)$ can be reproduced. This implies that nontrivial bounds can only be placed if $|c_{0}|<|x|$ and/or $|c_1|<|x||t|$.

\subsection{Non-classicality tests based on dimension witnesses}

Considering systems of a fixed dimension, quantum communication can outperform classical communication. This advantage can be revealed by using dimension witnesses. Specifically, by using a well-chosen quantum strategy involving states of Hilbert space dimension $d$, it is possible to violate certain classical dimension witnesses of dimension $d$. More formally, we say that a dimension witness with the following property
\begin{align}\label{claswit}
W={\bf w} \cdot {\bf p} \leq C_d < Q_d
\end{align}
can be used as non-classicality tests for systems of dimension $d$. Consider a set of data $ {\bf p}_Q$ such that  $W = {\bf w} \cdot {\bf p}_Q  >C_d$. This implies the use of genuinely quantum systems for reproducing $ {\bf p}_Q$, under the assumption that the experiment involves systems of dimension $d$. In Section \ref{case studies}, we will discuss several examples.

\subsection{Quantifying quantum advantage} \label{noisetolerance}

It is useful to quantify the advantage offered by quantum resources over classical ones. In the present context, several figures of merit can be considered. First, the amount of violation of a given dimension witness could be used, however this will generally depend on how the witness is expressed, and will not allow one to compare different witnesses. Hence, here we use the notion of noise tolerance, which has a more physical interpretation, and will allow us to compare various witnesses.

Consider a quantum experiment (with systems of dimension $d$) and its corresponding set of data ${\bf p}_Q$, which is found to violate a classical dimension witness, i.e. $W = {\bf w} \cdot {\bf p}_Q  >C_d$. The noise tolerance of the quantum point ${\bf p}_Q$ for this dimension witness is defined as the minimal fraction of white noise, $\eta$, such that the distribution
\begin{align}\label{noise}
{\bf p}_0=(1-\eta) {\bf p}_QÊ+ \eta {\bf p}_{\mathbb{I}}
\end{align}
does not violate the witness, i.e. $W = {\bf w} \cdot {\bf p}_0  =C_d$. Here ${\bf p}_{\mathbb{I}}$ denotes white noise, i.e. $p_{\mathbb{I}}(b,s|x,t,y)= \frac{1}{|b| \, |s|}$ is the uniform distribution for all $x,t,y$.

In a practical context, considering noisy distributions of the form \eqref{noise} is quite natural, due to unavoidable technical imperfections, e.g. losses or misalignment of the preparations.

\subsection{Bounded noise tolerance in prepare-and-measure scenarios involving qubits}

It turns out that the noise tolerance of qubit strategies is bounded for any dimension witness in the prepare-and-measure scenario. More precisely, any set of data obtained from qubits and projective measurements can be reproduced using one classical bit if the noise level $\eta$ satisfies
\begin{align}\label{noisemax}
\eta \geq \eta^{*} = 1-\frac{1}{k_{3}} \approx 0.34 ,
\end{align}
where $k_{3}$ is the Grothendieck constant \cite{Grot} of order three \footnote{Note that only upper and lower bounds are known for $k_{3}$; see e.g. T. V\'ertesi, Phys. Rev. A {\bf 78}, 032112 (2008).}. Hence, in the prepare-and-measure scenario, no dimension witness for classical bits and projective measurements can be violated for $\eta \geq \eta^{*}$.

We give a proof of the above statement. Consider that the choice of preparation is specified by a vector $\vec{x} \in \mathbb{R}^3$, which represents the Bloch vector of the desired qubit state. Similarly the measurement is specified by a Bloch vector $\vec{y} $, representing the observable $M_{\vec{y}} = \vec{y}\cdot \vec{\sigma}$ (with outcomes $b= \pm 1$), where $\vec{\sigma}= (\sigma_x,\sigma_y,\sigma_z)$ denotes the vector of Pauli matrices. The expected data is therefore
\be p(b|\vec{x},\vec{y}) = \frac{1+ b \, \vec{x}\cdot \vec{y}}{2} . \label{teleport}
\ee

Any such data can be reproduced classically by sending two bits \cite{toner}. In oder to see this, consider that the preparation and measurement devices share a singlet state $\ket{\psi^-}=\frac{1}{\sqrt{2}}\left(\ket{01}-\ket{10}\right)$. In order to prepare a qubit state corresponding to vector $\vec{x}$, measure the observable $\vec{x}\cdot \vec{\sigma}$ on (half of) the singlet. The result of this measurement is $a=\pm1$. Then, the state of the other half of the singlet (held by the measuring device) is given by the Bloch vector $-a\vec{x}$. By performing a measurement of the observable $-a\vec{y}\cdot \vec{\sigma}$ on this half of the state, we recover the data \eqref{teleport}. The protocol thus requires one bit of communication (to send $a$), and one singlet state. Using only classical resources, the protocol requires two bits of communication (as the simulation of the singlet state can be done with one bit of communication \cite{toner}).

Now, let us see what one can do using only a single bit of communication. The main point is that the simulation of a sufficiently noisy singlet state can be done without communication. That is, there exists a local hidden variable model (for projective measurements) for the state
\be \label{noisysing} \rho = w \ket{\psi^-}\bra{\psi^-} + (1-w)\mathbb{I}/4 \ee
for $w\leq 1/k_{3}\leq 0.66$ \cite{Acin}. Considering such a noisy singlet state in the above protocol, we see that it is possible to simulate the data \eqref{teleport} with probability $w$; with probability $1-w$ we obtain the distribution ${\bf p}_{\mathbb{I}}$. Hence, with a noise level  $\eta \geq \eta^*=1-\frac{1}{k_{3}}$, any qubit strategy can be simulated with one classical bit (and shared randomness).

As mentioned, the above result holds only if the measurement device performs a projective measurement. Since any two outcome qubit measurement can be written as a convex mixture of projective measurements, the result can be extended to all two outcome scenarios. One can extend further to general positive operator-valued measurements at the cost of a larger $\eta^*$ by using Werner's model \cite{werner} for the state \eqref{noisysing} with $w=\frac{1}{2}$, leading to $\eta^*=\frac{1}{2}$. This follows from the fact that Werner's model can be seen as a local hidden state model \cite{wisemansteering}, hence the model is valid if general measurements are performed on one side (the trusted party). 

\section{Case studies} \label{case studies}

We now present several case studies, illustrating the relevance of the concepts and tools discussed above. We first discuss two examples of networks of the form Fig. 2, where preparation, transformation, and measurement devices are `in a line'. We then discuss two examples based on a different network, featuring two separate preparations devices and one measurement device. Note that such a network has been considered in different contexts. Notably, this was studied in communication complexity, in the so-called simultaneous message passing model \cite{harry_review}, e.g. quantum fingerprinting \cite{fingerprint}, but also for the black-box certification of entangled measurements \cite{VN11,rabello,bennet}, and the Pusey-Barrett-Rudolph theorem \cite{PBR}.

In all cases quantum systems are shown to provide significant advantage over classical systems of the same dimension. Moreover, in all examples (except for the third one), this quantum advantage is stronger compared to the simple prepare-and-measure scenario, in terms of noise tolerance. This suggests that the simulation of quantum strategies becomes significantly harder in the case of networks, even if they feature only few devices.

\subsection{Three devices in a line: simple case}\label{ptm1}

We start with the network of Fig. \ref{line}, considering one of the simplest (non-trivial) configurations in terms of the number of inputs and outputs. Specifically, we have $|x|=3$ and $|t| = |y|= |b|=2$. Note that the transformation device does not give any outcome (i.e. $|s|=1$). We label the inputs and outputs: $x \in \{0,1,2\}$ and $t ,y,b \in \{0,1\}$. Hence a set of data is characterized by $D=24$ probabilities  $p(b|x,t,y)$. However, considering normalization conditions, this number is reduced to 12; specifically, the probabilities $p(1|x,t,y)=1-p(0|x,t,y)$ are redundant and can thus be omitted.

Applying the method described in Section \ref{geo} we have fully characterized the polytope $\mathbb{P}_2$, that is, the set of distributions achievable for $c_0,c_1 \in \{0,1\}$. Using the software \emph{PORTA}, we could find the complete list of facets of $\mathbb{P}_2$, which can be grouped (under relabeling of inputs and outputs) into 1870 inequivalent classes of dimension witnesses \footnote{For the full list of inequalities, contact \emph{joseph.bowles@unige.ch}}.

\begin{figure*}
\includegraphics[scale=0.707106781186547]{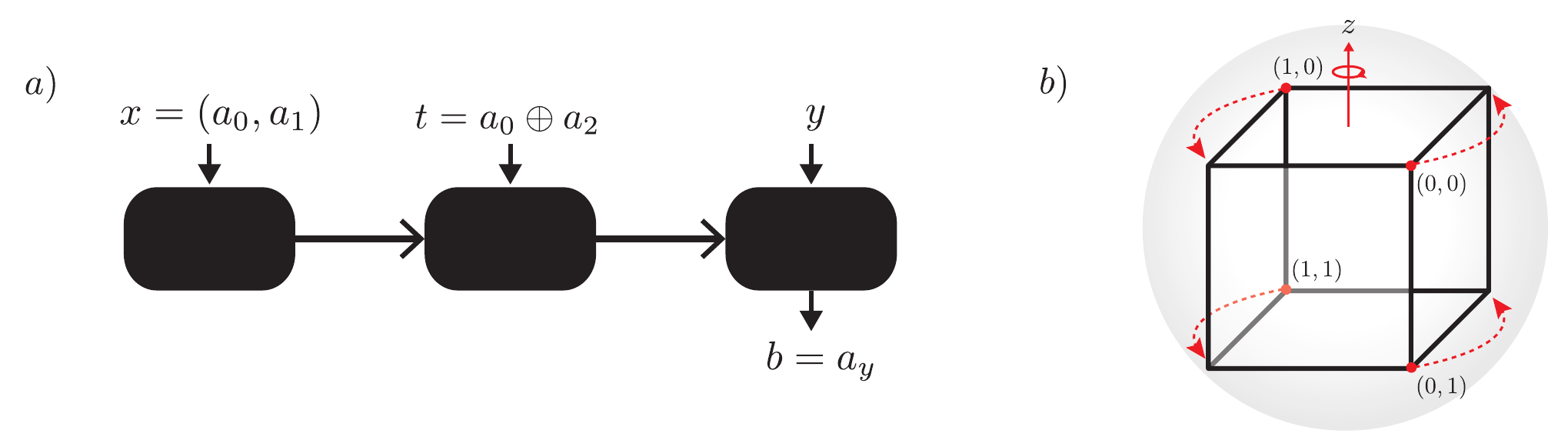}
\caption{\label{racfig} a) The network corresponding to the distributed $3 \rightarrow 1$ random-access-code (case study B). Three random bits $a_0,a_1,a_2$ are used to generate the inputs $x$ and $t$. Upon receiving input $y=0,1,2$, the measurement device should output $b=a_y$. The dimension witness $W_{D-RAC}$
(see Eq. \eqref{DRAC}) quantifies the average success probability.  b) Optimal qubit strategy. The four qubit preparations (red dots, corresponding to $(a_0,a_1)$) are given by the vertices of a cube inscribed inside the Bloch sphere. Upon receiving input $t=a_0\oplus a_2=1$, the transformation device performs a rotation of $\pi/2$ around the $z$ axis if $t=1$, and the identity otherwise. Finally, by performing a measurement in the $x,y,z$ directions, maximal information about $a_0,a_1,a_2$ (respectively) is obtained. }
\end{figure*}

Here, we present one class of tight dimension witnesses, a member of which can be written in simple form:
\begin{align}\label{wit1}
W_J = &p_{011}+p_{101}+p_{110}+p_{200}  \nonumber \\
&  -p_{000}-p_{001}-p_{010}-p_{211}\leq 2,
\end{align}
where we write $p_{xty}=p(b=0|x,t,y)$. A simple strategy using $c_0,c_1 \in \{ 0,1\}$ that reaches $W_J=2$ is as follows. The preparation devices sends $c_0=0$ for inputs $x=0,2$, but sends $c_{0}=1$ if $x=1$. Upon receiving $c_0$ and input $t$, the transformation device sends $c_1=c_0\oplus t$ to the measurement device (where $\oplus$ denotes addition modulo 2). Finally, the measurement device outputs $b=c_1\oplus y$. Note also that using classical trits, $c_0,c_1 \in \{ 0,1,2\}$, we can achieve $W_J = 4$, the maximal possible value.

Using qubits we can significantly outperform classical bits. Consider general pure qubit preparations:
\be \ket{\psi(\theta,\phi)}=\cos(\frac{\theta}{2})\ket{0}+ \sin(\frac{\theta}{2}) \exp(i\phi)\ket{1} .
\ee
Specifically, for preparations $x=0,1,2$ take $\ket{\psi(\frac{\pi}{2},0)}$, $\ket{\psi(\frac{\pi}{2}, \frac{3\pi}{4})}$ and  $\ket{\psi(\frac{\pi}{2}, \frac{-3\pi}{4})}$ respectively. Next consider the transformation device, parametrized by
\be \Phi_{t=0} = \mathbb{I}_2   \quad , \quad \Phi_{t=1} = \exp(-i\frac{\pi}{4} \sigma_z), \ee
where $\sigma_z=\text{diag}(1,-1)$ is the Pauli $z$ matrix. Finally, for the measuring device, we have the measurement operators
\be M_{0|0} &=&   \ket{\psi(\frac{\pi}{2}, \frac{-3\pi}{4})} \bra{ \psi( \frac{\pi}{2},\frac{-3\pi}{4}) } \\
M_{0|1} &=&  \ket{\psi(\frac{\pi}{2}, \frac{3\pi}{4})} \bra{ \psi(\frac{\pi}{2},\frac{3\pi}{4}) }.
\ee
Calculating the resulting probabilities, via eq. \eqref{qprobdet}, and inserting them into eq. \eqref{wit1}, we obtain
\be W_J = 2+\sqrt{2} \approx 3.41. \ee
The above qubit strategy thus clearly violates the witness \eqref{wit1}, and can therefore not be reproduced with classical bits; classical trits must be used. Numerical optimization strongly suggests that this qubit strategy is optimal.

The noise tolerance of the above qubit strategy is
\be \eta = \sqrt{2}-1 \approx 0.41  . \ee
Notably, this value exceeds the bound $\eta^*\approx 0.34$ (see Section \ref{noisetolerance}) for any prepare-and-measure scenario.
Hence the advantage offered by qubits compared to classical bits is stronger compared to what is possible in the prepare-and-measure scenario.

\subsection{Distributed $3\rightarrow1$ random access code}\label{ptm2}

As a second example, we consider a task inspired from the information-theoretic task of a random access code (RAC) \cite{Ambainis}.

Specifically, we consider a distributed version of the $3\rightarrow1$ RAC featuring three devices in a line (see Fig. \ref{racfig} (a)).
Consider 3 bits $a_0,a_1,a_2$ randomly taken from a uniform distribution. These bits will determine the inputs of the preparation and transformation devices, namely: $x=(a_0,a_1)$ and $t=a_0\oplus a_2$. Again, the transformation device has no output. The measuring devices has a ternary input $y=0,1,2$. Similarly to a RAC, the goal is to have the output $b=a_y$. Hence we can define the following witness (for the scenario $|x|=4$, $|t|=|b|=2$, $|y|=3$, and $|s|=1$) which is the average success probability:
\small
\begin{align}\nonumber  \label{DRAC}
W_{\scalebox{0.5}{D-RAC}}= \frac{1}{24}\;\sum_{\mathclap{\substack{a_0a_1 \\a_2y}}}\; p (b=a_{y}|x=(a_0,a_1),t=(a_0\oplus a_2),y) \leq C_{d}.
\end{align}
\normalsize

We first discuss the case of classical communication. For bits we obtain the bound $C_{2}=\frac{2}{3}$, which can be achieved as follows. The preparation device sends $c_0=a_0$ to the transformation device, who in turn sends $c_1=c_0$ to the measurement device for both inputs $t=0,1$. The measurement device outputs $b=c_1=a_0$. Hence, for $y=0$ we always have $b=a_{y}$. However for $y=1,2$, success is only achieved with probability $1/2$. Overall, this leads to $C_{2}=\frac{2}{3}$. For the case of classical trits, $c_0,c_1 \in \{ 0,1,2\}$, we get $C_3=19/24$. In order to achieve success with probability one, i.e. $W_{\scalebox{0.5}{D-RAC}}=1$, eight-dimensional systems are required.

Next, we discuss quantum strategies. Using qubits, we can achieve up to
\be W_{\scalebox{0.5}{D-RAC}}=Q_2 = \frac{1}{2}(1+\frac{1}{\sqrt{3}})\approx0.79. \ee
The optimal strategy is the following. For input $x=(a_0,a_1)$, choose preparations
\begin{align}
\ket{\psi((-1)^{a_1}\arccos(\frac{1}{\sqrt{3}})+\pi a_{1},\frac{\pi}{4}+\pi a_0)},
\end{align}
which lie at four of the vertices of the cube inscribed inside the Bloch sphere (see Fig. \ref{racfig} (b)). The transformations are given by:
\begin{align}
\Phi_{t=0}=\mathbb{I}_{2}  \quad , \quad \Phi_{t=1}=\exp(i \frac{\pi}{4} \sigma_{z}).
\end{align}
Finally, the measuring device performs a measurement in one of three mutually unbiased bases:
\begin{align}
M_{0|0}&\;\;=\;\;\ket{\psi( \frac{\pi}{2},0)} \bra{ \psi( \frac{\pi}{2},0) } &= \;\; \ket{+x} \bra{+x}&  \nonumber \\
M_{0|1}&\;\;=\;\;\ket{\psi(0,0)} \bra{ \psi(0,0) } &= \;\; \ket{+z} \bra{+z}&  \nonumber \\
M_{0|2}&\;\;=\;\;\ket{\psi(\frac{\pi}{2},\frac{\pi}{2})} \bra{ \psi(\frac{\pi}{2},\frac{\pi}{2}) } &= \;\; \ket{+y} \bra{+y}&.
\end{align}

The noise tolerance of this strategy is given by
\be \eta=1-\frac{1}{\sqrt{3}}\approx 0.43 \ee
which again exceeds the bound for the prepare-and-measure scenario, $\eta^*\approx 0.34$.

Finally, let us comment on the relation of the above game and the standard (prepare-and-measure) $3\rightarrow 1$ RAC. We first note that the optimal qubit strategies for $W_{\scalebox{0.5}{D-RAC}}$ and the standard RAC are in fact essentially the same \cite{Laura}. Specifically, the qubit states arriving at the measuring device are identical in both cases (given inputs $(a_0,a_1,a_2)$). Hence, this qubit is unaffected by the fact that the inputs are now distributed between the preparation and transformation devices. Indeed, the ability of implementing unitary transformations is central here.

Interestingly, the situation is very different for the case of classical bits. While the average probability of success is $3/4$ in the standard RAC, the fact that the inputs are now distributed decreases the average score to $2/3$. The reason for this that the optimal strategy in the standard RAC is to send $c=\text{maj}(a_0,a_1,a_2)$, where $\text{maj}(.)$ denotes the majority function. However using this strategy requires access to all the input bits $a_0,a_1,a_2$, which none of the devices in distributed RAC has. The consequence of this is that the noise tolerance of qubit strategies is enhanced in the distributed version of the game, as we showed above.

\subsection{Two preparation devices, one measurement device: simple case}

We now consider a scenario with two preparation devices sending communication to a measurement device (see Fig. \ref{nldcfig} (a)). A simple non-trival scenario here is one in which both preparation devices receive a ternary input. We denote the input of the first device $x_0 \in \{0,1,2\}$, and the input of the second $x_1 \in \{0,1,2\}$. The measurement device has no input (i.e. a fixed measurement) and provides a binary output $b=\{0,1\}$. That is, we have $|x_0|=|x_1|=3$, $|y|=1$ and $|b|=2$.

We consider the case in which the channels carry classical bits, i.e. $c_0,c_1 \in \{0,1\}$. In this case we have fully characterized the polytope $\mathbb{P}_2$: it features 13 non-trivial classes of facets which we present in Appendix \ref{appendixA}. Here we focus on one particular class (witness 1 in appendix), represented by the following witness:
\small
\begin{align}\label{entwit}
W_K=-p_{00}+p_{01}+p_{02}-p_{10}-p_{12}+p_{20}+p_{21}-p_{22}\leq 2 ,\nonumber
\end{align}
\normalsize
where $p_{x_0 x_1}=p(b=0|x_0, x_1)$. An optimal classical bit strategy is as follows. The first preparation device sends $c_0=0$ for $x_0=0,1$ and $c_0=1$ for $x_{0}=2$. The second preparation device sends $c_1=1$ for $x_{1}=0,1$ and $c_1=0$ for $x_{1}=2$. The measurement device then outputs $b=c_0\cdot c_1 \oplus 1$. Clearly, sending classical trits achieves the maximum $W_K = 4$.

Let us now discuss strategies involving qubits. Via numerical optimization we expect a maximal quantum violation of
\be W_K =Q_2 =  \frac{5}{2} .\ee
This can be achieved using the following strategy. The two preparation devices prepare the same states, i.e. we have $\rho_{x_1}=\rho_{x_2}$ for $x_1=x_2$. For inputs $x_1=x_2=0,1,2$, the preparations are
\begin{align}
\ket{\psi(-\alpha,0)},\quad  \ket{0} , \quad \ket{\psi(\alpha,0)}
\end{align}
respectively, with $\alpha=2\arccos \sqrt{\frac{3}{8}}$. The measurement operator for outcome $b=0$ is a projection onto the entangled subspace:
\begin{align}
M_{0}=\ket{\phi^-}\bra{\phi^{-}}+\ket{\xi(\gamma)}\bra{\xi(\gamma)},
\end{align}
with $\gamma=\arccos \sqrt{\frac{1}{10}}$ and where
\begin{align}
\ket{\xi(\gamma)}=\cos \gamma \ket{01} - \sin \gamma \ket{10}.
\end{align}
The corresponding noise tolerance is $\eta=0.2$.

It is relevant to consider a situation in which one channel sends a qubit, while the other one sends a classical bit. Performing numerical optimization, we find a maximal value of $W_{K}\approx 2.337$ for this case.

Finally, one may also ask if this witness could be used to detect entangled measurements, similarly to Ref. \cite{VN11}. Specifically, one can derive an upper bound on $W_K$ for separable measurement operators of the form $M_{b}=\sum_{i} M_{b,1}^{i}\otimes M_{b,2}^{i}$ where $M_{b,k}^i$ is a positive operator acting on the system sent by preparation device $k$. Numerical tests suggest that the optimal value is $W_{K}\approx 2.337$. Hence we find the same value as for the above case of hybrid qubit/bit channels. Therefore, we expect that a value $W_{K}>2.337$ certifies that (i) both channels send qubits and (ii) the measurement is non-separable, i.e. has (at least) one entangled eigenstate. Note that the witness \eqref{entwit} has been discussed before in \cite{migueldimension} in a similar context, where upper bounds of  $W_{K}\approx2.506$ and $W_{K}\approx2.377$ were found for the case of general and unentangled measurements, supporting our findings. 

\subsection{Nonlocal dense coding}\label{case4}

As the last example, we present a dimension witness for a task which can be viewed as a nonlocal version of dense coding \cite{wiesner}. As in the previous example, we consider the case of two preparation devices and one measuring device.

\begin{figure}
\includegraphics[scale=0.6]{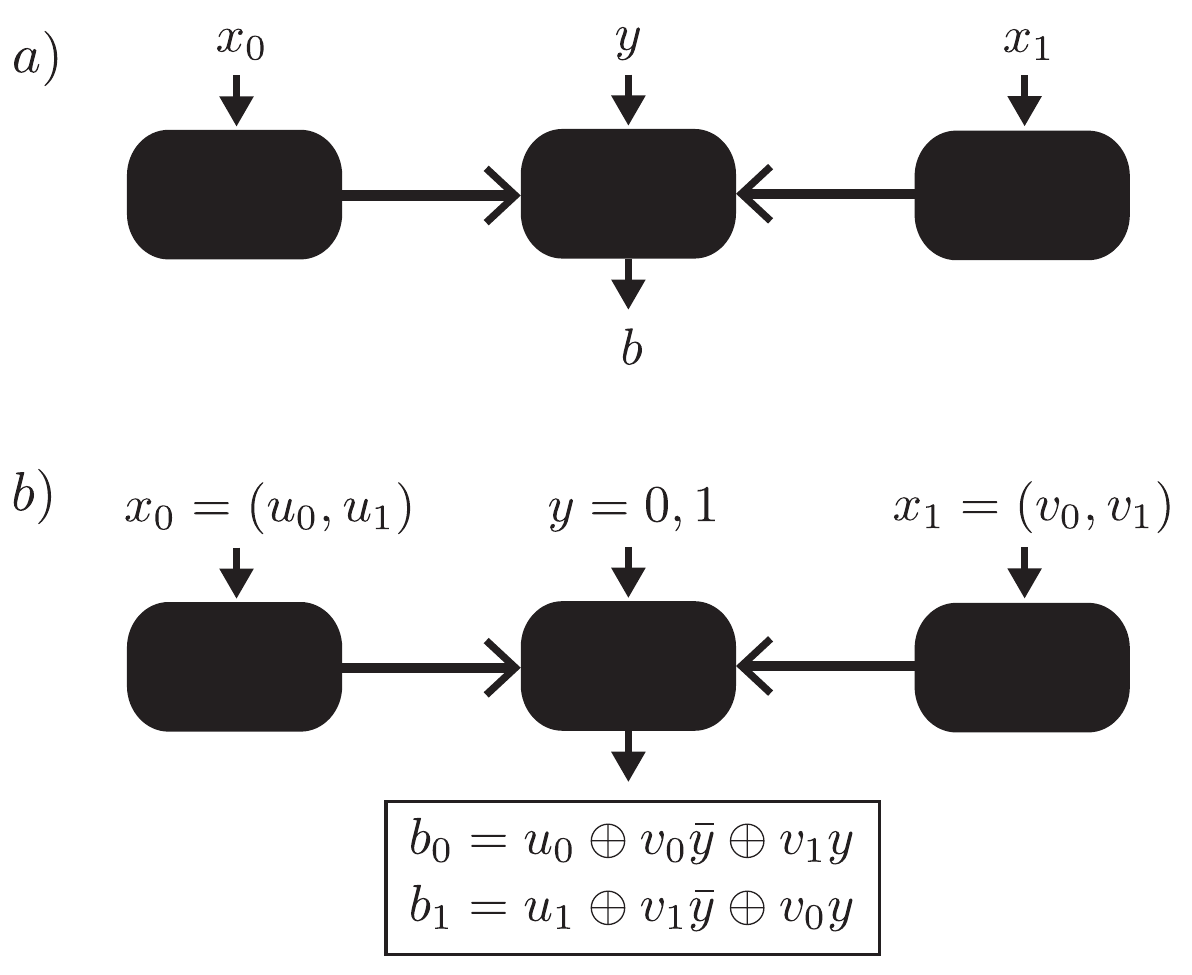}
\caption{\label{nldcfig} (a) A simple network involving two preparation devices (left and right) and a single measurement device (center). (b) In case study D, we discuss a dimension witness for this network, referred to as nonlocal dense coding.}
\end{figure}

Here each preparation device receives two input bits: $x_0=(u_0,u_1)$ for the first and $x_1=(v_0,v_1)$ for the second. The measurement device receives $y=0,1$ as input, and provides two output bits ${\bf{b}}=(b_0,b_1)$. The rules of the game are the following (see Fig. \ref{nldcfig}(b)). On the one hand, for $y=0$, the outputs should satisfy $(b_0,b_1)=(u_0\oplus v_0,u_1\oplus v_1)$. On the other hand, for $y=1$, the output bits should satisfy $(b_0,b_1)=(u_0\oplus v_1,u_1\oplus v_0)$. Furthermore, there is a penalty if both $b_0$ and $b_1$ are guessed incorrectly. This corresponds to the witness
\small
\begin{align}\label{nldcwit}
W_{D}=&\langle (b_0,b_1)=(u_0\oplus v_0\bar{y} \oplus v_1 y , u_1\oplus v_1\bar{y} \oplus v_0 y )\rangle \\
&-\langle (\bar{b}_0,\bar{b}_1)=(u_0\oplus v_0\bar{y} \oplus v_1 y , u_1\oplus v_1\bar{y} \oplus v_0 y )\rangle \leq C_d\nonumber,
\end{align}
\normalsize
where $\bar{y}=y \oplus 1$, and the average $\langle \cdot \rangle$ is taken over all inputs:
\begin{align}
\langle (b_0,b_1) \rangle = \frac{1}{32}\;\sum_{\mathclap{\substack{u_0,u_1\\ v_0,v_1, y}}}p(b_0,b_1|u_0,u_1,v_0,v_1,y).
\end{align}

%

Let us discuss the case of classical communication. For bits, we have $C_{2}= \frac{1}{4}$ which can be achieved as follows. The first preparation devices sends communication $c_0=u_0\cdot u_1$. Similarly, the second device sends $c_1=v_0\cdot v_1$. The measurement device then outputs $(b_0,b_1)=(c_0\oplus c_1, c_0 \oplus c_1)$. Using classical trits, we get $C_3=\frac{9}{16}$. Indeed, sending four dimensional systems achieves success probability one.

Next, consider qubit strategies (see Appendix \ref{nldcap} for more details). Here we can achieve
\be W_D=Q_2 = \frac{1}{2} \ee
which appears optimal from numerical tests. This corresponds to a noise tolerance of $\eta=\frac{1}{2}$, which represents a considerable improvement over the simple prepare-and-measure scenario. The strategy is the following. The preparation devices send qubit states
\begin{align}
&\sigma_x^{u_1}\sigma_z^{u_0}\ket{\psi(\frac{\pi}{4},0)}, \\
&\sigma_x^{v_1}\sigma_z^{v_0}\ket{\psi(-\frac{3\pi}{4},0)}
\end{align}
for the first and second preparation devices respectively. The measurement device then performs a projective measurement onto the entangled basis
\begin{align} \label{bell}
M_{b_0b_1|y}=\ket{\phi(b_0,b_1,y)}\bra{\phi(b_0,b_1,y)},
\end{align}
where
\begin{align}
\ket{\phi(b_0,b_1,y)}=\sigma_x^{b_1}\sigma_z^{b_0}\otimes H^y \ket{\psi^{-}},
\end{align}
$\ket{\psi^{-}}=\frac{1}{\sqrt{2}}(\ket{01}-\ket{10})$ is the singlet state and $H=\frac{1}{\sqrt{2}}$\scalebox{0.7}{$\begin{pmatrix}1&1 \\ 1 & -1 \end{pmatrix}$} is the Hadamard matrix. Note that by using qutrits, one can reach $Q_{3}\approx0.598$ according to numerical optimization. Hence we obtain the following relations $C_2 < Q_2 < C_3 < Q_3$.

Additionally, one may also wish to consider the possibility that the devices share quantum correlations (i.e. initial entanglement). Allowing for this considerably enhances the success probability (still using qubit communication), which becomes maximal, that is $W_{D}=1$. The strategy is the following. The preparation devices now share a singlet state. Upon receiving the inputs $x_0=(u_0,u_1)$ and $x_1=(v_0,v_1)$, the preparation devices locally rotate the singlet state to
\begin{align}
(\sigma_x^{u_1}\sigma_z^{u_0}) \otimes (\sigma_x^{v_1}\sigma_z^{v_0}) \ket{\psi^-}.
\end{align}
The measurement device performs the same measurement as above (see \eqref{bell}). The noise tolerance for this strategy is $\eta=\frac{3}{4}$.

\section{Discussion} \label{discussion}

We have discussed the problem of testing the dimension and non-classicality in communication networks. We have presented methods for addressing these problems, generalizing the concept of dimension witnesses to networks, and discussed several illustrative examples.

We believe our results raise several natural questions. Firstly, it would be interesting to investigate the separation between classical and quantum dimension in more general networks. In particular, what is the classical communication cost (i.e. how many classical dimensions are required) for simulating qubit networks? A potential direction for tackling this problem would be to find a family of dimension witnesses for a scenario featuring one preparation device and one measurement device, but any number of transformation devices in between (here we gave examples for the case of a single transformation device). Notably, Galv\~ao and Hardy \cite{galvao} proved that, in the case of an infinite number of transformation devices, classical systems of infinite dimension are required for simulating a single qubit. The game discussed in \cite{galvao} can be recast as a dimension witness. Proving a similar result for the case of a finite number of transformation devices would be relevant. Going beyond qubits is also interesting. In fact, for quantum systems of dimension $d \geq 3$, it is not known whether an exact simulation is possible with classical systems of finite dimension, even in the simplest prepare-and-measure scenario.


From a more applied perspective, the ideas discussed could find applications in quantum information processing. Recent works discussed protocols for which the security is based on dimension witnesses, so-called semi-device-independent protocols \cite{PB11,Letal11,HWL12, YCL, Woodhead}. For instance, quantum key distribution and randomness expansion can be achieved, assuming only that the devices prepare and measure qubit systems. Moving to more general networks may allow for more robust and efficient protocols, and other information-theoretic tasks.

\section{Acknowledgments}

This work is supported by FNP programme TEAM and NCN through grant 2014/14/E/ST2/00020, and the Swiss National Science Foundation (grant PP00P2\_138917 and Starting grant DIAQ), and SEFRI (COST action MP1006).

\begin{appendix}

\section{All dimension witnesses for a simple network} \label{appendixA}

Here we present all dimension witnesses for the scenario of Fig. \ref{nldcfig} (a) with $|x_0|=|x_1|=3$, $|y|=1$ and $|b|=2$. In this scenario, considering classical communication $c_0,c_1 \in \{ 0,1 \}$, there exist 13 non-trivial facets (i.e. facets that do not correspond to the normalization of probabilities). We present the witnesses in tabular form, using the notation

\begin{align}
\begin{pmatrix}
w_{00} & w_{01} & w_{02} \\
w_{10} & w_{11} & w_{12} \\
w_{20} & w_{21} & w_{22}
\end{pmatrix}
\leq C_2
\end{align}
\newline
to describe the witness
\begin{align}
\sum_{x_0=0}^{2}\sum_{x_1=0}^{2}w_{x_0x_1}p(0|x_0,x_1)\leq C_2.
\end{align}
\newline
The 13 witnesses are:
\newline
\begin{align}
&1) \;\begin{pmatrix}
-1 & 1 & 1 \\
-1 & 0 & -1 \\
1 & 1 & -1
\end{pmatrix}
\leq 2 \nonumber
&&2) \;\begin{pmatrix}
2 & -1 & 1 \\
2 & 0 & -2 \\
0 & -1 & 1
\end{pmatrix}
\leq 4 \nonumber
\\[8pt]
&3)\; \begin{pmatrix}
1 & -1 & 1 \\
1 & -2 & -3 \\
0 & 2 & -2
\end{pmatrix}
\leq 2 \nonumber
&&4) \;\begin{pmatrix}
1 & 1 & 0 \\
1 & -1 & 0 \\
0 & 1 & -1
\end{pmatrix}
\leq 3 \nonumber
\\[8pt]
&5)\; \begin{pmatrix}
2 & 2 & 0 \\
1 & -2 & 0 \\
-1 & 1 & -1
\end{pmatrix}
\leq 4 \nonumber
&&6)\; \begin{pmatrix}
1 & -2 & 3 \\
2 & 0 & -2 \\
-1 & 2 & 1
\end{pmatrix}
\leq 6 \nonumber
\\[8pt]
&7)\;\begin{pmatrix}
1 & -1 & 2 \\
1 & 0 & -1 \\
-1 & 1 & 1
\end{pmatrix}
\leq 4 \nonumber
&&8)\; \begin{pmatrix}
1 & -1& 2 \\
2 & 0 & -2 \\
-1 & 1 & 0
\end{pmatrix}
\leq 4 \nonumber
\\[8pt]
&9)\; \begin{pmatrix}
2 & -2 & 4 \\
4 & -1 & -5 \\
-2 & 1 & -1
\end{pmatrix}
\leq 6 \nonumber
&&10)\; \begin{pmatrix}
1 & -1& 2 \\
2 & -3 & -5 \\
-1 & 3 & -3
\end{pmatrix}
\leq 3 \nonumber
\\[8pt]
&11)\; \begin{pmatrix}
2 & 2 & -1 \\
1 & -1 & 0 \\
-1 & 1 & 0
\end{pmatrix}
\leq 4
\quad\;
&&12)\; \begin{pmatrix}
1 & -2 & 3 \\
3 & 1 & -2 \\
-2 & 3 & 1
\end{pmatrix}
\leq 8 \nonumber
\\[8pt]
&13)\; \begin{pmatrix}
1 & -1 & 1 \\
1 & 0 & -1 \\
0 & 0 & 0
\end{pmatrix}
\leq 2 .
\nonumber
\end{align}

Note that the last witness (13) is in fact a lifting from the simplest prepare-and-measure scenario featuring 3 preparations and two binary measurements. This can be seen by imagining that the first preparation device in our scenario simply acts as a classical input for the measurement device, i.e. $x_1$ takes the role of $y$ in the prepare-and-measure scenario. Since the channel supports bits, then we must have $y=0,1$. In the final witness we see that $x_1=0,1$ corresponds to $y=0,1$ and $x_{1}=2$ is never used (since we have all zeros on the bottom row of the witness). Upon interpreting $x_1$ as $y$ in a prepare-and-measure scenario, the final witness then corresponds to Equation 6 of \cite{DimWit}.\newline

\section{Quantum violation in nonlocal dense coding}\label{nldcap}
Here we calculate explicitly the values of \eqref{nldcwit} for strategies using qubits. We first consider the case where the devices do not share initial entanglement. To ease notation we define
\begin{align}
\ket{h_{+}}=\ket{\psi(\frac{\pi}{4},0)} \quad;\quad \ket{h_{-}}=\ket{\psi(-\frac{3\pi}{4},0)}.
\end{align}

Following the preparations and measurements given in the main text, we have
\small
\begin{align}
&p(b_0,b_1|u_0,u_1,v_0,v_1,y) \nonumber \\
&=|\bra{\psi^-}\sigma_x^{b_1\oplus u_1}\sigma_{z}^{b_0\oplus u_0}\otimes H^y\sigma_x^{v_1}\sigma_{z}^{v_0}\ket{h_+}\ket{h_-}|^2 \nonumber \\
&=|\bra{\psi^-}\sigma_x^{b_1\oplus u_1\oplus v_1 \bar{y} \oplus v_0 y}\sigma_{z}^{b_0\oplus u_0 \oplus v_0 \bar{y} \oplus v_1 y}\otimes \mathbb{I}\ket{h_+}\ket{h_-}|^2,
\end{align}
\normalsize
where in the last line we have used
\begin{align}
H\sigma_x^{v_1}\sigma_z^{v_0}=\sigma_z^{v_0}\sigma_x^{v_1}H
\end{align}
and
\begin{align}
H\ket{h_{\pm}}=\pm \ket{h_{\pm}}.
\end{align}
By writing $\ket{\psi^-}=\frac{1}{\sqrt{2}}(\ket{h_+}\ket{h_-}-\ket{h_-}\ket{h_+})$ we see that the probability that $(b_0,b_1)=(u_0 \oplus v_0 \bar{y} \oplus v_1 y,u_1 \oplus v_1 \bar{y} \oplus v_0 y)$ is given by
\begin{equation}
|\bra{\psi^-}\ket{h_+}\ket{h_-}|^{2}=\frac{1}{2}.
\end{equation}
The probability that both bits are guessed incorrectly, i.e. $(\bar{b}_0,\bar{b}_1)=(u_0 \oplus v_0 \bar{y} \oplus v_1 y,u_1 \oplus v_1 \bar{y} \oplus v_0 y)$ is
\begin{equation}
|\bra{\psi^-}\sigma_{x}\sigma_{z}\otimes\mathbb{I}\ket{h_+}\ket{h_-}|^{2}=0.
\end{equation}
Hence, we achieve $W_{D}=\frac{1}{2}$. In order to treat the case in which the preparation devices share entanglement, we need to replace the state $\ket{h_+}\ket{h_-}$ by the singlet state $\ket{\psi^-}$. Hence the probability that $(b_0,b_1)=(u_0 \oplus v_0 \bar{y} \oplus v_1 y,u_1 \oplus v_1 \bar{y} \oplus v_0 y)$ becomes
\begin{equation}
|\braket{\psi^-|\psi^{-}}|^{2}=1
\end{equation}
and the game is won perfectly.

\end{appendix}

\end{document}